\documentclass[11pt]{article}
\usepackage[dvips]{graphicx}
\usepackage{latexsym}
\usepackage{amssymb}

\newcommand{\C}{\mathbb{C}}
\newcommand{\CP}{\mathbb{CP}}
\newcommand{\RP}{\mathbb{RP}}

\newcommand{\R}{\mathbb{R}}

\renewcommand{\d}{\mathrm{d}}

\def\be{\begin{equation}}
\def\ee{\end{equation}}

\def\Sm{\Sigma}

\def\OO{\cal O}

\def\ov{\overline}
\def\g{\mathfrak{g}}
\def\l{{\lambda}}
\def\p{\partial}
\def\ov{\overline}

\def\ll{\lambda}

\def\OO{{\cal O}}

\begin{document}
\pagestyle{plain}
\title{\vskip -70pt
\begin{flushright}
{\normalsize DAMTP-2007-8}\\
\end{flushright}
\vskip 80pt
{\bf Multidimensional integrable systems and deformations of Lie 
algebra homomorphisms}\vskip 20pt}

\author{Maciej\ Dunajski
\thanks{email 
{\tt m.dunajski@damtp.cam.ac.uk}}\\[10pt]
{\sl Department of Applied Mathematics and Theoretical Physics} \\[3pt]
{\sl University of Cambridge} \\[3pt]
{\sl Wilberforce Road, Cambridge CB3 0WA, U.K.} \\[10pt]
James~D.E.\ Grant\thanks{email 
{\tt james.grant@univie.ac.at}}\\[10pt]
{\sl Fakult{\"a}t f{\"u}r Mathematik, Universit{\"a}t Wien}\\[3pt]
{\sl Nordbergstrasse 15, 1090 Wien, Austria}\\[10pt]
and
\\[10pt]
Ian A.B.\ Strachan\thanks{email
{\tt i.strachan@maths.gla.ac.uk}}
\\[10pt]
{\sl Department of Mathematics}\\[3pt]
{\sl University of Glasgow}\\[3pt]
{\sl Glasgow G12 8QW, U.K.}\\[15pt]
}
\date{}
\maketitle
\thispagestyle{empty}
\begin{abstract}
We use deformations of Lie algebra homomorphisms to
construct deformations of dispersionless integrable systems
arising as symmetry reductions of anti--self--dual Yang--Mills
equations with a gauge group Diff$(S^1)$.
\end{abstract}
\newpage

\section{Introduction}
\setcounter{equation}{0}

A dispersionless limit of PDEs is taken by rescaling the independent 
variables $X^a\longrightarrow X^a/\varepsilon$, and taking the limit
$\varepsilon \longrightarrow 0$. This is a delicate procedure,
as the limit of the solutions of a given PDE does not usually correspond
to  solutions of the limiting dispersionless equation.
Moreover, inequivalent PDEs may have the 
same dispersionless limit, so the problem
\begin{itemize}
\item
Recover the original PDE from its dispersionless limit
\end{itemize}
is, of course, ill posed. Some progress can nevertheless be made
if the dispersionless equation is integrable, and one insists that
its dispersive analogue is also integrable. In the next section we shall 
explain how dispersionless limits of solitonic PDEs are equivalent
to the WKB quasi-classical approximation of the associated linear problems.
This suggests that the reconstruction of the dispersive solitonic system
should involve a quantisation of some kind. 
 
Such a quantisation procedure has been developed in the seminal work of
Kupershmidt~\cite{Ku}. This procedure is based on the Moyal product, and
works well if the Lie algebra underlying the dispersionless linear problem
is the algebra sdiff$(\Sigma^2)$ of divergence-free vector fields on a 2-surface $\Sigma^2$. 
This is the case for the dKP and $SU(\infty)$
Toda equations in $2+1$ dimensions. Similar progress can also be made 
in higher 
dimensions and, indeed, one of us has constructed integrable deformations of Plebanski's first heavenly equation~\cite{S92} by replacing the underlying Poisson bracket with the Moyal bracket.

The idea of deforming integrable systems while retaining
the integrability of the resulting equation has now been studied from
a number of different points of view:

\begin{itemize}

\item Takasaki studied properties of the deformed heavenly equations  and described
how solutions may be described in terms of a Riemann-Hilbert splitting in a
Moyal algebra valued loop group~\cite{Ta92}. Extensions of this led to Moyal-KP hierarchies~\cite{Ta94}
and deformations of the self-dual Yang-Mills equations~\cite{Ta01}.
The deformed Riemann-Hilbert procedure was recently fully developed 
by Formanski and Przanowski~\cite{F_P1, F_P2}.

\item Nekrasov and Schwarz introduced instantons on 
noncommutative
space-time~\cite{NS}. This led to the development of noncommutative soliton
equations. These may be viewed as a deformation of the standard, commutative,
soliton equations. Many of these may be studied as reductions of the 
noncommutative
self-dual Yang-Mills equations~\cite{Ham, Lench}.

\item Associated to any Frobenius manifold is a hierarchy of
integrable equations of hydrodynamic type. Integrable deformations of 
these equations
arise naturally when one studies the genus expansion in the corresponding
topological quantum field theories~\cite{DZ}.

\end{itemize}

In the present  paper deformations of multidimensional integrable systems are 
based on the algebra\footnote{In the remainder of this paper
the superscript, denoting the dimension of the manifold, will be dropped.}
diff$(\Sigma)$, the Lie algebra of vector fields on $\Sigma$, where $\Sigma\cong S^1 {\rm~or~}{\R}$.
It turns out, however, that this algebra admits no non-trivial deformations \cite{Lich}.
However an alternative method of deforming these integrable systems
may be developed. This method is based on the approach of 
Ovsienko and Rogers~\cite{OR} 
where a homomorphism from
diff$(\Sigma)$ to the Poisson algebra on $T^*\Sigma$ can be used to construct
non-trivial deformations. We shall use this idea to construct integrable
deformations of various equations 
associated to the algebra diff$(\Sigma)\,.$

It should be pointed out that the original, undeformed, equations have a 
natural
interpretation in terms of twistor theory, via the non-linear graviton 
construction
and its variants. It would seem desirable to develop a \lq 
deformed\rq~version of
twistor theory that would encode solutions of the deformed equations as some
sort of deformed holomorphic conditions. This idea was what was behind the
paper~\cite{S97}, but the problem remains open (though see~\cite{M} for
some ideas on how Nijenhuis structures may be deformed).

\section{Dispersionless limit in $2+1$ dimensions}
\setcounter{equation}{0}

Certain dispersionless integrable systems can arise from solitonic systems
in a following way: Let
\begin{eqnarray*}
A\Big(\frac{\p~}{\p X}\Big)&=&
\frac{\p^n~}{\p X^n}+a_1(X^a)\frac{\p^{n-1}~}{\p X^{n-1}}+\dots+a_n(X^a), \\
B\Big(\frac{\p~}{\p X}\Big)&=&\frac{\p^m~}{\p X^m}
+b_1(X^a)\frac{\p^{m-1}~}{\p X^{m-1}}+\dots+b_m(X^a)
\end{eqnarray*}
be differential operators on $\R$ with coefficients depending on local
coordinates $X^a=(X, Y, T)$ on $\R^3$. The overdetermined linear system
\[
\Psi_Y=A\Big(\frac{\p~}{\p X}\Big)\Psi, \qquad 
\Psi_T=B\Big(\frac{\p~}{\p X}\Big)\Psi
\]
admits a solution $\Psi(X, Y, T)$ on a
neighbourhood of initial point $(X, Y_0, T_0)$ for arbitrary initial data
$\Psi(X, Y_0, T_0) = f(X)$ if and only if the integrability conditions
$\Psi_{YT}=\Psi_{TY}$, or
\be
\label{soleq1}
A_T-B_Y+[A, B]=0
\ee
are satisfied.
The nonlinear system (\ref{soleq1}) for $a_1, \dots, a_n, b_1, \dots, b_m$
can be solved by the inverse scattering transform (IST).
Integrable systems which admit a Lax representation
(\ref{soleq1}) will be referred to as solitonic, or dispersive.

The dispersionless limit~\cite{Z94} is obtained by substituting
\[
\frac{\p}{\p X^a}=\varepsilon\frac{\p}{\p x^a}, \qquad
\Psi(X^a)=\exp{(\psi(x^a/\varepsilon))},
\]
and taking the limit $\varepsilon\longrightarrow 0$.
In the limit the commutators of differential operators are replaced by
the Poisson brackets of their symbols according to the relation
\[
\frac{\p^k}{\p X^k}\Psi\longrightarrow (\psi_x)^k\Psi, \qquad
[A, B]\longrightarrow
\frac{\p A}{\p\l}\frac{\p B}{\p x}-\frac{\p A}{\p x}\frac{\p B}{\p \l}
=\{A, B\}, \qquad \l=\psi_x,
\]
where $A, B$ are polynomials in $\l$, with coefficients depending on
$x^a=(x, y, t)$. The dispersionless limit of the system
(\ref{soleq1}) is
\be
\label{bezdys}
A_t-B_y+\{A, B\}=0.
\ee
Nonlinear differential equations of the form (\ref{bezdys})
are called dispersionless integrable systems.
One motivation for studying of dispersionless integrable systems is
their role in constructing partition functions in topological field
theories~\cite{Kri94}.

A natural approach to solving (\ref{bezdys}) would be an attempt to
take a quasi-classical limit of the IST which linearises (\ref{soleq1}).
This does not yield the expected result, as the quasi-classical limit of
the Lax representation for (\ref{soleq1}) is the system of Hamilton--Jacobi
equations
\[
\psi_y=A(\psi_x, x^a), \qquad \psi_t=B(\psi_x, x^a),
\]
with \lq two times\rq\ $t$ and $y$, and the initial value problem for
(\ref{bezdys}) would require a reconstruction of a potential from
the asymptotic form of the Hamiltonians.
This classical inverse scattering problem is so far open.

There are alternative methods of solving (\ref{bezdys})
\cite{Kon3, FK04, Ta90, DMT00}.
In particular the mini-twistor approach of~\cite{DMT00}
works as follows:
The system (\ref{bezdys}) is equivalent to the integrability $[L, M]=0$ 
of a two-dimensional distribution of vector fields
\be
\label{BDcong}
L=\p_t-B_{\l}\p_x+B_x\p_{\l}, \qquad M=\p_y-A_{\l}\p_x+A_x\p_{\l}
\ee
on $\R^3\times\RP^1$.
Assume that $L, M$ are real analytic, and complexify $\R^3$ to $\C^3$.
The mini-twistor space $Z$ is the two complex dimensional
quotient manifold
\[
Z= \C^3\times \CP^1/(L, M), \qquad \l\in \CP^1,
x^a\in \C^3.
\]
That is to say that the local coordinates on $Z$ lift to functions on
$\C^3\times \CP^1$ constant along $L, M$.

The mini-twistor space is equipped with a three parameter family
of certain rational curves. All solutions to (\ref{bezdys})
can in principle be reconstructed from a complex structure of
the mini-twistor space.

In fact the twistor approach outlined above is capable of solving a 
wider class
of equations. We shall therefore generalise the notion of the dispersionless
integrable systems by allowing distributions of vector fields more general
than (\ref{BDcong}). The derivatives $A_\l, A_x, B_\l, B_x$ of the symbols
$(A, B)$ of operators can be replaced by
independent polynomials $A_1, A_2, B_1, B_2$ in $\l$ with coefficients
depending on $(x, y, t)$
\be
\label{BDcong1}
L=\p_t-B_1\p_x+B_2\p_{\l}, \qquad M=\p_y-A_1\p_x+A_2\p_{\l}.
\ee
If $A_1, B_1$ are linear in $\l$ and $A_2, B_2$ are
at most cubic in $\ll$ then the rational curves in $Z$ have normal bundle
$\OO(2)$ (the line bundle over $\CP^1$ with transition functions
$\ll^{-2}$ from the set $\ll\neq\infty$ to $\ll\neq 0$ i.e.
Chern class $2$) and the three--dimensional
moduli space of such curves in $Z$ can be parametrised by $(x, y, t)$.
Allowing polynomials of higher degrees would lead to
hierarchies of dispersionless equations.
We take the integrability of this generalised distribution
(\ref{BDcong1}) as our definition of the dispersionless integrable system.
The definition is intrinsic in a sense that it does not refer to an
underlying dispersive equation.

\section{Diff$(S^1)$ dispersionless integrable systems}
\setcounter{equation}{0}

In this section two integrable systems associated with the gauge group
Diff$(S^1)$ will be given. The first has been extensively
studied in~\cite{Pa03, FK04, D04, MaSh02, DS05, Man_San06, OR2}, so only 
a new gauge theoretic
description will be given - the reader is referred to these earlier
papers for more details. The second system, which arises from a Nahm-type
system, is new and this system is discussed in more detail.

\subsection{A $(2+1)$ dimensional dispersionless integrable system}
An example of a dispersionless system which is integrable in the sense
of the outlined twistor correspondence is given by the following
distribution
\be
\label{Lax11}
L=\p_t-w\p_x-\ll\p_y, \qquad
M=\p_y+u\p_x-\ll\p_x.
\ee
A linear combination of this distribution leads
to a special case of (\ref{BDcong1}) with $A_2=B_2=0$.
Its integrability leads to the pair of quasi-linear PDEs
\be
\label{PMA}
u_t+w_y+uw_x-wu_x=0, \qquad u_y+w_x=0,
\ee
for two real functions $u=u(x, y, t), w=w(x, y, t)$.
This system of equations has recently been studied in
\cite{Pa03, FK04, D04, MaSh02, DS05, Man_San06, OR2} 
in connection with Einstein--Weyl
geometry, hydrodynamic chains and symmetry reductions of anti--self--dual
Yang--Mills equations. From the twistor point of view (\ref{PMA})
is invariantly characterised \cite{D04} by requiring that the mini--twistor space $Z$
fibres holomorphically over $\CP^1$.
The second equation can be used to introduce a potential $H$ such that
$u=H_x, w=-H_y$. The first equation then gives
\be
\label{Heq}
H_{xt}-H_{yy}+H_{y}H_{xx}-H_{x}H_{xy}=0.
\ee

The system (\ref{PMA}) arises as a symmetry reduction
of the anti--self--dual Yang Mills equations in signature $(2, 2)$
with the infinite--dimensional gauge
group Diff$(\Sm)$ and two commuting translational symmetries
exactly one of which is null~\cite{DS05}. This combined with the
embedding of $SU(1, 1)\subset$ Diff$(\Sm)$ gives rise
to explicit solutions to (\ref{PMA}) in terms of solutions to
the nonlinear Schr\"{o}dinger equation, and
the Korteweg de Vries equation~\cite{DS05}.

The Lie algebra of the group of diffeomorphisms
Diff$(\Sigma)$, where $\Sigma=S^1$ or $\R$,
is isomorphic to the infinite--dimensional
Lie algebra of functions on $\Sigma$ with the Wronskian
\be
\label{wron}
<f, g>:=fg_x-f_xg
\ee
as the Lie bracket, where $f, g\in C^{\infty}(\Sigma)$, and $x$ is a local
coordinate on $\Sigma$. An alternative gauge--theoretic interpretation can be given to
(\ref{PMA}):
Observe that the first equation in (\ref{PMA}) can be interpreted
as the flatness of a gauge connection on $\R^2$, where the gauge group is
Diff$(\Sigma)$.
Indeed, choose local coordinates $(t, y)$ on $\R^2$ and consider
${\cal A}\in\Lambda^1(\R^2)\otimes \C^{\infty}(\Sm)$ of the general form
\[
{\cal A}=-w\d t+u\d y,
\]
where 
$u, w: \mathbb{R}^2 \rightarrow C^{\infty}(\Sm)$
depend on $(x, y, t)$.

The flatness of this connection yields
\[
\d {\cal A}+{\cal A}\wedge {\cal A}=(u_t+w_y+<u, w>)\d t\wedge\d y=0,
\]
as claimed. Therefore the connection is a pure gauge, and can be written
as ${\cal A}=g^{-1}\d g$, where $g = g(x, y, t) \in \mathrm{Map}(\R^2, \mathrm{Diff}(\Sm))$, and
\[
w=-g^{-1}g_t, \qquad u=g^{-1}g_y.
\]
The second equation in (\ref{PMA}) yields the following system
\be
(g^{-1}g_y)_y-(g^{-1}g_t)_x=0,
\ee
where $g=\exp{({\cal A})}$
is a finite diffeomorphism of $\Sm$, and terms like $g^{-1}g_t$
should be understood as
\[
g^{-1}g_t={\cal A}_t-<{\cal A}, {\cal A}_t>+\frac{1}{2}<{\cal A},
<{\cal A}, {\cal A}_t>>+\dots\;.
\]
\subsection{A $(3+1)$ dimensional dispersionless integrable system}

In this section we shall present another example of an integrable
system associated to the Lie algebra of Diff$(S^1)$. We shall first
write it as a Nahm system
\be\label{GS1}
\dot{{\mathbf{e}}}_i=\frac{1}{2}\varepsilon_{ijk}[{\mathbf{e}}_j,
{\mathbf{e}}_k],\qquad i = 1, 2, 3
\ee
where ${\mathbf{e}}_i$ are vector fields on an open set in $\R^4$ given by
\[
{\mathbf{e}}_i=\frac{\p}{\p y^i}-N_i(x, y^j)\frac{\p}{\p x},
\]
and $(x, y^j )$ are local coordinates.
Rewrite
(\ref{GS1}) as
\be
\label{GS2}
\p_x{N}_i+\varepsilon_{ijk}\p_j N_k-\frac{1}{2}\varepsilon_{ijk}<N_j, N_k>=0.
\ee
We shall now discuss the origin
and possible applications of (\ref{GS2})
\begin{enumerate}
\item
Any solution to (\ref{GS2}) defines a hyperHermitian conformal
structure represented by the metric
\be
\label{conf_str}
{g} = {\bf{n}}^2 + \delta_{ij} \d y^i\d y^j ,
\ee
where
\[
{\bf{n}}=\d x+N_i\d y^i.
\]
The three complex structures ${\bf I}_i, i=1, 2, 3$ 
satisfying the algebra of quaternions 
\[
{\bf I}_i\,{\bf I}_j=-\delta_{ij}{\bf 1}+\varepsilon_{ijk}{\bf I}_k
\]
are given by
\[
{\bf I}_i({\bf{n}})= \d y^i.
\]
These formulae  together with the algebraic relations satisfied by
${\bf I}_j$ determine the complex structures uniquelly, e.g.
\[
{\bf I}_i( \d y^j)=-\delta_{ij}({\bf{n}})
+\varepsilon_{ijk} \d y^k.
\]
One way
to impose integrability of the complex structures is to use the explicit
form of the complex structures
on the basis $(\d y^1, \d y^2, \d y^3, {\bf{n}})$ and demand
that the space $\Lambda^{(1, 0)}$ is closed under exterior 
differentiation. We begin by defining a basis of self-dual $2$-forms
\[
{\bf{\Sigma}}^i = {\bf{n}} \wedge \d y^i +
\frac{1}{2}\varepsilon^{ijk}\d y^j \wedge \d y^k.
\]

The integrability of the complex structures is then equivalent to
the anti-self-duality of the two-form $\d{\bf{n}}$:
\begin{equation}
{\bf{\Sigma}}^i \wedge \d {\bf{n}} = 0.
\label{int}
\end{equation}
This condition is equivalent to (\ref{GS2}).

A dual formulation leads to the Lax pair of vector fields, which
is a special form of the hyperHermitian Lax pair~\cite{D99, GS, C01}.
To see it set $\mathbf{e}_4=\p_x$ and define complex vector fields
\[
\mathbf{w} = \mathbf{e}_1 - i \mathbf{e}_2, \qquad \mathbf{z} =
\mathbf{e}_3 - i\mathbf{e}_4.
\]
The system (\ref{GS2}) is equivalent to the commutativity of the Lax pair
\be
\label{laxhc}
\left[
\mathbf{w} - \lambda {\overline{\mathbf{z}}}, \mathbf{z} + \lambda 
{\overline{\mathbf{w}}} \right] = 0
\ee
for all values of the parameter $\ll$.
\item
The system {(\ref{GS2})} arises as a symmetry reduction
of the anti--self--dual Yang Mills equations on $\R^4$
with the infinite--dimensional gauge
group Diff$(S^1)$ and one translational symmetry. In fact any such
symmetry reduction is gauge equivalent to
(\ref{GS2}). To see it consider the flat metric on $\R^4$ which
in double null coordinates $w=y^1+iy^2, z=y^3+i y^4$ takes the
form
\[
\d s^2=\d z\d \ov{z}+\d w\d \ov{w},
\]
and choose the volume element
$\d w\wedge\d \ov{w}\wedge\d z\wedge\d \ov{z}$.
Let $A\in T^*\R^4\otimes\g$ be a connection one--form, and
let $F$ be its curvature two--form. Here $\g$ is the Lie algebra of some
(possibly infinite dimensional) gauge group $G$.
In a local trivialisation $A=A_\mu\d y^\mu$
and $F=(1/2)F_{\mu\nu}\d y^{\mu}\wedge\d y^{\nu}$, where
$F_{\mu\nu}=[D_{\mu}, D_{\nu}]$
takes its values in $\g$. Here
$D_\mu=\p_\mu+A_\mu$ is the covariant derivative.
The connection is defined up to gauge transformations
$A\rightarrow b^{-1}Ab-b^{-1}\d b$, where $b\in \mbox{Map}(\R^4, G)$.
The ASDYM equations on $A_\mu$ are $F=-\ast F$, or
\[
F_{wz}=0, \qquad F_{w\ov{w}}+F_{z\ov{z}}=0,\qquad
F_{\ov{w}\ov{z}}=0.
\]
These equations are equivalent to the commutativity of the Lax pair
\be
\label{LaxSDYM}
L=D_w-\ll D_{\ov{z}}, \qquad M=D_z+\ll D_{\ov{w}}
\ee
for every value of the parameter $\ll$.

We shall require that the connection possesses a symmetry which in
our coordinates is given by $\p/\p y^4$.
Choose a gauge such that the Higgs field $A_{4}$ is a constant
in $\g$. Now choose $G=\mbox{Diff}(S^1)$, so that the components
of the one form $A$ become vector fields
on $S^1$. We can choose a local coordinate $x$ on $S^1$ such that
$A_{4}=\p_x$, and $A_i=-N_i\p_x$,
where $N_i=N_i(x, y^j)$ are smooth functions on $\R^4$.
The Lax pair (\ref{LaxSDYM}) is identical to (\ref{laxhc})
and the ASDYM equations
reduce to the first order PDEs (\ref{GS2}).

\item Example.  An ansatz ${\bf N}(x, y^j)= f(x){\bf A}(y^j)$,
where $\mathbf{N}=(N_1, N_2, N_3)^T$,
reduces (\ref{GS2}) to a pair of linear equations
\[
\dot{f}=c f, \qquad c{\bf A}+\nabla \wedge {\bf A}=0,
\]
where $c$ is a constant. If $c = 0$ then ${\bf N}$ may be absorbed into a redefinition of the coordinate $x$ in the metric~(\ref{conf_str}). Therefore we assume $c \neq 0$. 
We set $c=1$ by rescaling $y^j$ and
solve for $f=\exp{(x)}$ reabsorbing another constant of integration into 
${\bf A}$.  Now define a new coordinate
$\rho =\exp{(-x)}$. Rescaling the  metric (\ref{conf_str}) yields
\be
\label{beltrami_met}
\hat{g}= \rho \,\d{\bf y}^2+\rho^{-1}\,(\d \rho -{\bf A}\cdot \d {\bf y})^2.
\ee
This metric is hyper--hermitian iff the vector ${\bf A}(y^j)$ satisfies the Beltrami equation
\be
\label{beltrami}
{\bf A} + \nabla \wedge {\bf A} = 0.
\ee
This is a slight improvement of the result of \cite{W_P} where it is claimed
that (\ref{beltrami_met}) is ASD iff (\ref{beltrami}) holds. 

The Beltrami equation implies that ${\bf A}$ is divergence--free
and satisfies
$
\triangle{\bf A}+{\bf A}=0 ,
$
where $\triangle=\nabla^2$ is the scalar Laplacian on $\R^3$ acting on components of
${\bf A}$.
Existence of solutions of equation~(\ref{beltrami}), at least in
the analytic case, 
can be proved by an application of 
the Cartan-K\"{a}hler theorem 
(c.f. Example~3.7 in Chapter~III of~\cite{BCGGG}). 

\item The system (\ref{GS2}) can be put in the hydrodynamic form
\[
\p_x{\mathbf{N}}=\mathbf{M}\; \mathrm{curl}\,{\mathbf N},
\]
where
\[
{\mathbf M}=-\left(\begin{array}{rrr}
1 & N_3 & -N_2 \\
-N_3 & 1 & N_1 \\
N_2 & - N_1 & 1
\end{array}\right)^{-1}\,.
\]

A different analytic continuation of (\ref{GS2})  can be obtained 
at the level of the hyperHermitian
geometry. This comes down to looking for conformal structures
(\ref{conf_str}) of signature $+ + - -$. To achieve this we regard
$y_1, y_2, N_1, N_2$ as imaginary, and define
\[
Y_1=iy_1, \quad Y_2=iy_2, \quad Y_3=y_3, \qquad {\cal N}_1=iN_1,
\quad {\cal N}_2=iN_2,\quad {\cal N}_3=N_3.
\]
The desired system for ${\cal N}_i={\cal N}_i(Y^j, x)$ arises from
(\ref{GS2}).
\end{enumerate}
Clearly there are many further properties of these dispersionless systems that may be
studied. We now turn our attention to the construction of non-trivial
integrable deformations of equations (\ref{PMA}) and (\ref{GS2}).

\section{Dispersive deformations}
\setcounter{equation}{0}

Given a dispersionless integrable system it is natural to ask whether
it arises as a limit of some dispersive (or solitonic) system.
One would expect the reconstruction of a solitonic system to
involve a quantisation of some kind, because taking a dispersionless
limit of (\ref{soleq1}) was equivalent to a quasi-classical limit of the wave
function $\Psi(X^a)$.
This is indeed the case, and the paradigm example is provided by
the connection
between the Kadomtsev--Petviashvili (KP) equation
and its dispersionless limit dKP.
One can reconstruct KP
from dKP by expressing the latter
in the form (\ref{bezdys}), and replacing the
Poisson brackets by the Moyal bracket~\cite{Ku, Ta90, S95, S97, PF99}.
The infinite series involved in a Moyal product truncates in this case, 
because
the symbols
$A$ and $B$ are polynomials in momentum $\l$.
The deformation parameter can then be set to one, and
removed from the construction. It seems however that this beautiful example
is rather exceptional, and that the reconstruction of
dispersive systems (if at all possible)
is in general non-unique
and can lead to systems which involve a formal power series.

Generalising the definition of the dispersionless systems to
non--Hamiltonian distributions
$(L, M)$ like (\ref{Lax11}) makes things even worse,
as the Poisson bracket is not present, and
the connection with known quantisation procedures of a classical phase--space
has been lost.
It could be argued that (\ref{Heq})
should be regarded as its own deformation as it admits a dual 
(classical, and quantum) description:
It is a solitonic system (\ref{soleq1}) with
\[
A=H_X\frac{\p}{\p X}, \qquad B=H_Y\frac{\p}{\p X}
\]
or a dispersionless limit (\ref{bezdys}) with $A=\l H_x, B=\l H_y$.

One attempt to find a dispersive analogue of (\ref{PMA}) would be to 
use the centrally extended Virasoro algebra in place of diff$(\Sigma)$. 
Recall that such a procedure has been used to produce dispersive 
systems from dispersionless systems in a different context. 
Namely~\cite{OK}, one can view the periodic Monge equation $u_t = u 
u_x$ as the equation for affinely parametrised geodesics with respect 
to the right-invariant metric on Diff$(S^1)$ constructed from the $L^2$ 
inner product on the Lie algebra. Going to the central extension, one 
finds that affinely parametrised geodesics on the Virasoro-Bott group 
correspond to solutions of the KdV equation. (For a recent review of 
such constructions, see~\cite{M06}.) In the current situation, we view 
a general element of the extended algebra as a pair
\[
(f, a):=f(x)\; \frac{\d}{\d x}-iac
\]
where $a\in\R$ does not depend on $x$, and
$c$ is a constant.
Assuming that $x$ is a periodic variable
the modified commutation relation are
\[
[f\p_x, g\p_x]_c=<f, g>\p_x+\frac{ic}{48\pi}\int (f_{xxx}g-fg_{xxx})\d x,
\]
and we see that the central term is a function of $(t, y)$ only. Applying
this procedure to (\ref{Heq}) with
\be
\label{H_vira}
H(x, y, t)=\sum_k h_k(y, t)L^k,
\ee
where $L^k$ are generators of the centrally extended Virasoro algebra
satisfying
\be
\label{vira}
[L^k, L^m]=(k-m)L^{k+m}+\frac{c}{12}k(k^2-1)\delta_{k, -m}
\ee
would modify only one equation in the infinite chain of PDEs for the functions
$h_k$.

In the remaining part of this paper we shall
present a construction\footnote{An alternative approach which we have not explored
would be to
consider a quantum deformation of the Virasoro algebra as in~\cite{skao95}
and its free boson realisation
\be
\label{q_vira}
[T_m, T_n]=-\sum_{l=1}^{\infty}f_l(T_{n-l}T_{m+l}-T_{m-l}T_{n+l})
-\frac{(1-q)(1-t^{-1})}{1-s}(s^n-s^{-m})\delta_{m+n, 0},
\ee
where $s=qt^{-1}$, and coefficients $f_l$ are given by
\[
f(z)=\sum_{l=0}^\infty f_lz^l=\exp{\Big(\sum_{n=1}^\infty
\frac{1}{n}\frac{(1-q^n)(1-t^{-n})}{1+s^n}z^n\Big)}.
\]
The ordinary Virasoro algebra (\ref{vira})
is recovered as $q\rightarrow 1$.
Applying (\ref{q_vira}) to (\ref{H_vira}) would lead to a q-deformed
analog of (\ref{Heq}).}
which leads to non--trivial
dispersive analogues of (\ref{PMA}),
or its equivalent form (\ref{Heq}), and (\ref{GS2}).

\subsection{Deforming Lie algebra homomorphisms}
\label{DLAH}
To find a non-trivial deformation one would wish to deform the
Lie algebra of vector fields on $\Sigma$, but this algebra is
known not to admit any non-trivial deformations~\cite{Lich}.

We shall choose a different route~\cite{OR},
and deform the standard homomorphism between diff$(\Sm)$ and the
Poisson algebra on $T^*\Sm$, the point being that the
homomorphisms between Lie algebras
can admit non--trivial deformations even if one of the algebras
is rigid. A deformation of (\ref{Heq}) is achieved in two steps, each
introducing a parameter. In the first step we shall deform the
embedding of diff$(\Sm)$ into the Lie algebra of volume--preserving
vector fields on $T^*\Sm$. This introduces the first parameter $\mu$.
The second step is a deformation quantisation of the
first one: The Poisson algebra on $T^*\Sm$ is the quasi-classical limit
of the Lie algebra of pseudo--differential operators on $\Sm$, so
(working at the level of symbols) one quantises the deformed
homomorphism by using the deformed associative product of symbols of
pseudo--differential operators rather than a pointwise commutative
product of functions. This introduces the second parameter $\varepsilon$.
In what follows we shall be interested in the polynomial deformations
rather than the formal ones.

The standard embedding $\pi:\mbox{vect}(\Sm)\rightarrow
C^{\infty}(T^*\Sm)$ is given by contracting a vector field
$X_f=f(x)\p_x$ with a canonical one--form $\Theta$ on $T^*\Sm$. 
In our case $T^*\Sm=\R\times \Sm$ and the Lie algebras 
algebras $C^{\infty}(S^1)$ (with the Wronskian bracket) and
$\mbox{vect}(S^1)$ (with the  Lie bracket) are isomorphic so we can regard
$\pi$ as defined on  $C^{\infty}(\Sigma)$.

If
$\l$ is a local coordinate on the fibres of $T^*\Sm$, and $\Theta=\l\d x$,
the map $\pi$ is
explicitly given by
\[
(\pi(f))(\ll, x):=\ll f(x).
\]
It is a Lie algebra homomorphism as
\[
\{\pi(f), \pi(g)\}=\pi(<f, g>).
\]
Given $\mu\in \R$ define~\cite{OR}
\begin{eqnarray*}
({\pi}_{\mu}(f))(\ll, x) &=& (\pi(f))(\ll, x+\mu/\l) = \ll f(x+\mu/\l)\\
&=&
\l \Big(f(x)+f'(x)\frac{\mu}{\ll}+\frac{1}{2!}f''(x)
\Big(\frac{\mu}{\ll}\Big)^2+\dots\;\Big).
\end{eqnarray*}
Note that $\{\pi_{\mu}(f), \pi_{\mu}(g)\}=\pi_{\mu}(<f, g>)$, so that $\pi_{\mu}$ is also a Lie algebra
homomorphism between diff$(\Sigma)$ and 
sdiff$(T^*\Sigma)$.

The next step is motivated by the canonical quantisation
$\l \rightarrow \p/\p x$.
For any functions $F, G \in C^{\infty}(T^*\Sm)$ which are also
allowed to depend on a parameter $\mu$ define
the Kupershmidt-Manin product
\be
\label{Moyalproduct}
F\star G=\sum_{k=0}^\infty \frac{\epsilon^k}{k!}\frac{\p^k F}{\p \l^k}
\frac{\p^k G}{\p x^k}\,.
\ee
(this is equivalent, under an $\epsilon$-valued change of variable
to the Moyal product)
and set
\be
\{F, G\}_{\varepsilon}=\frac{1}{\epsilon}(F\star G-G\star F).
\ee
The Poisson bracket is recovered in the limiting procedure
\[
\lim_{\epsilon\rightarrow 0}\;\{F, G\}_{\epsilon}=\{F, G\},
\]
but the deformed bracket is equal to the Poisson bracket
for all $\epsilon$ if $F, G$ are linear in $\l$. This is why the
first deformation parameter $\mu$ is needed.

We are now ready to
propose the dispersive analog of the dispersionless equation
(\ref{Heq}) and (\ref{GS2}).

\begin{enumerate}

\item Let $\hat{H}(\ll, x, y, t ;\mu, \epsilon)=\pi_\mu(H)$ take
values in an algebra of formal power series in $\epsilon$ with
an associative product defined by (\ref{Moyalproduct}). The deformed
analogue of equations (\ref{Heq}) is:
\be
\label{Omeq}
\hat{H}_{xt}-\hat{H}_{yy}-\{\hat{H}_x, \hat{H}_y\}_{\epsilon}=0.
\ee

\item Let $\hat{N_i}(\ll,x^i;\mu, \epsilon)=\pi_\mu(N_i)$ take
values in an algebra of formal power series in $\epsilon$ with
an associative product defined by (\ref{Moyalproduct}). The deformed
analogue of equation (\ref{GS2}) is:
\be
\label{GS2deformed}
\p_x{\hat{N}}_i+\varepsilon_{ijk}\p_j {\hat 
N}_k-\frac{1}{2}\varepsilon_{ijk}\{ {\hat{N}_j}, 
{\hat{N}_k}\}_{\epsilon}=0.
\ee
\end{enumerate}

\noindent Given a solution $\hat{H}$ of (\ref{Omeq}) such that
$(\mu\p_\mu-\ll\p_{\ll})(\ll^{-1}\hat{H})=0$, and $\ll^{-1}\hat{H}$
is smooth in $(\mu/\ll)$
we can construct
$H(x, y, t)$ satisfying (\ref{Heq}) by taking
any of the two limits $\mu\longrightarrow 0,
\epsilon\longrightarrow 0$ and similar remarks hold for equation 
(\ref{GS2deformed}).
Conversely, formal powers series (in the deformation parameters) 
solution may be constructed
from a solution to the original, undeformed, equation in an analogous 
manner to the way developed in~\cite{S92}. 
The extent to which such formal series converge in a suitable space of functions is as yet, however, unclear.

These deformed equations formally retain their integrability; the various manipulations hold at the
level of the Lax pair as well as at the level of the equations 
themselves. However, as remarked in the introduction, a direct 
twistor theory correspondence
for these equations is lacking, though one should be able to adopt the 
methods developed by
Takasaki~\cite{Ta92} and Formanski--Przanowski~\cite{F_P1, F_P2} to study 
the geometry of the corresponding  Riemann-Hilbert problem (in some
suitable Moyal algebra valued loop group).

\section*{Acknowledgements}
We thank Guido Carlet and Maciej Przanowski for useful discussions. MD 
is a member
of the European network Methods of Integrable Systems, Geometry and 
Applied Mathematics (MISGAM).
The work of JDEG was funded by START-project Y237--N13 of
the {Austrian Science Fund}. This paper was finished when MD and JDEG attended
the 07frg138 BIRS
workshop  {\em $\Xi$--Transform} held in Banff in March 2007.


\end{document}